\documentclass[a4paper,11pt]{article}
\pdfoutput=1 

\usepackage{jheppub} 

\usepackage[T1]{fontenc} 

\usepackage{amsmath}
\usepackage{graphicx,slashed,xcolor,multirow,bbold,mathtools,sidecap,tikz,bm,enumitem,booktabs,array}

\usepackage[normalem]{ulem}

\usepackage{tikz}
\usetikzlibrary{positioning,arrows}
\usetikzlibrary{decorations.pathmorphing}
\usetikzlibrary{decorations.markings}

\usepackage{subcaption}

\usepackage[compat=1.1.0]{tikz-feynman}
\newcommand{\beq}{\begin{equation}}
\newcommand{\eeq}{\end{equation}}
\newcommand{\bea}{\begin{eqnarray}}
\newcommand{\eea}{\end{eqnarray}}
\newcommand{\barr}{\begin{array}}
\newcommand{\earr}{\end{array}}

\usepackage{color}
\usepackage{colortbl}

\long\def\/*#1*/{}
\usepackage{color}
 \usepackage[normalem]{ulem}
\definecolor{darkgreen}{cmyk}{1,0,1,0.4}
\definecolor{darkred}{cmyk}{0,1,1,0.4}

\definecolor{lime}{HTML}{A6CE39}
\DeclareRobustCommand{\orcidicon}{\hspace{-2.1mm}
\begin{tikzpicture}
\draw[lime, fill=lime] (0,0) circle [radius=0.13] node[white] {{\fontfamily{qag}\selectfont \tiny \,ID}}; \draw[white, fill=white] (-0.0525,0.095) circle [radius=0.007]; 
\end{tikzpicture} \hspace{-3.2mm} }
\foreach \x in {A, ..., Z}{\expandafter\xdef\csname orcid\x\endcsname{\noexpand\href{https://orcid.org/\csname orcidauthor\x\endcsname} {\noexpand\orcidicon}}}

\title{Integrating Physics Inspired Features with Graph Convolution}

\author[]{Rameswar Sahu\orcidA{}}
\affiliation[]{\footnotesize Institute of Physics, Bhubaneswar, Sachivalaya Marg, Sainik School Post, Bhubaneswar 751005, India}
\affiliation[]{\footnotesize Homi Bhabha National Institute, Training School Complex, Anushakti Nagar, Mumbai 400094, India}
\emailAdd{rameswar.s@iopb.res.in}

\abstract{With the advent of advanced machine learning techniques, boosted object tagging has witnessed significant progress. In this article, we take this field further by introducing novel architectural modifications compatible with a wide array of Graph Neural Network (GNN) architectures. Our approach advocates for integrating capsule layers, replacing the conventional decoding blocks in standard GNNs. These capsules are a group of neurons with vector activations. The orientation of these vectors represents important properties of the objects under study, with their magnitude characterizing whether the object under study belongs to the class represented by the capsule. Moreover, capsule networks incorporate a regularization by reconstruction mechanism, facilitating the seamless integration of expert-designed high-level features into the analysis. We have studied the usefulness of our architecture with the LorentzNet architecture for quark-gluon tagging. Here, we have replaced the decoding block of LorentzNet with a capsulated decoding block and have called the resulting architecture CapsLorentzNet. Our new architecture can enhance the performance of LorentzNet by 20 \% for the quark-gluon tagging task.}


\begin{document} 

\tikzset{
  vector/.style={decorate, decoration={snake,amplitude=.4mm,segment length=2mm,post length=1mm}, draw},
  tes/.style={draw=black,postaction={decorate},
    decoration={snake,markings,mark=at position .55 with {\arrow[draw=black]{>}}}},
  provector/.style={decorate, decoration={snake,amplitude=2.5pt}, draw},
  antivector/.style={decorate, decoration={snake,amplitude=-2.5pt}, draw},
  fermion/.style={draw=black, postaction={decorate},decoration={markings,mark=at position .55 with {\arrow[draw=blue]{>}}}},
  fermionbar/.style={draw=black, postaction={decorate},
    decoration={markings,mark=at position .55 with {\arrow[draw=black]{<}}}},
  fermionnoarrow/.style={draw=black},
  scalar/.style={dashed,draw=black, postaction={decorate},decoration={markings,mark=at position .55 with {\arrow[draw=blue]{>}}}},
  scalarbar/.style={dashed,draw=black, postaction={decorate},decoration={marking,mark=at position .55 with {\arrow[draw=black]{<}}}},
  scalarnoarrow/.style={dashed,draw=black},
  electron/.style={draw=black, postaction={decorate},decoration={markings,mark=at position .55 with {\arrow[draw=black]{>}}}},
  bigvector/.style={decorate, decoration={snake,amplitude=4pt}, draw},
  particle/.style={thick,draw=blue, postaction={decorate},
    decoration={markings,mark=at position .5 with {\arrow[blue]{triangle 45}}}},
  gluon/.style={decorate, draw=black,
    decoration={coil,aspect=0.3,segment length=3pt,amplitude=3pt}}
}

\maketitle
\flushbottom

\section{Introduction}
\label{sec:intro}
Machine Learning techniques have been proven highly effective in many collider studies such as jet classification \cite{Cogan:2014oua,deOliveira:2015xxd,Baldi:2016fql,Barnard:2016qma,Komiske:2016rsd,ATL-PHYS-PUB-2017-017,Kasieczka:2017nvn,Bhimji:2017qvb,Macaluso:2018tck,Guo:2018hbv,Dreyer:2018nbf,Guest:2016iqz,Louppe:2017ipp,Cheng:2017rdo,Egan:2017ojy,Fraser:2018ieu,Almeida:2015jua,Pearkes:2017hku,Butter:2017cot,Roxlo:2018adx,Datta:2017rhs,Aguilar-Saavedra:2017rzt,Luo:2017ncs,Moore:2018lsr,Datta:2017lxt,Komiske:2017aww}, pileup removal \cite{Komiske:2017ubm}, constraining Effective Field Theories \cite{Brehmer:2018kdj,Brehmer:2018eca,DHondt:2018cww}, and several other analysis \cite{Collins:2018epr,DAgnolo:2018cun,DeSimone:2018efk,Hajer:2018kqm,Farina:2018fyg,Heimel:2018mkt,deOliveira:2017pjk,Paganini:2017hrr,deOliveira:2017rwa,Paganini:2017dwg,Andreassen:2018apy,Baldi:2014kfa,Baldi:2014pta,Searcy:2015apa,Santos:2016kno,Barberio:2017ngd,Duarte:2018ite,Abdughani:2018wrw,Lin:2018cin,Lai:2018ixk}. (See ref \cite{Larkoski:2017jix,Guest:2018yhq,Albertsson:2018maf,Radovic:2018dip} for some excellent reviews). Among these, the most popular among the high-energy community is the tagging of boosted objects (t, W, Z, H, q, g). Boosted objects are special in that it is usually impossible to identify the individual decay products of these objects in a collider environment. It is, therefore, suitable to reconstruct these objects with a larger radius jet (fat jet), with a radius chosen according to the particle's boost so that the object's cumulative properties can be studied. The field of boosted object tagging is further divided into several sub-fields based on the particle that originates the fat jet. Some examples include the separation of hadronically decaying top/W/Z/H jets from quark/gluon-initiated jets, separating $W^{+}$ fat jets from $W^{-}$ fat jets, separating W jets from Z jets, and distinguishing quark-initiated jets from gluon-initiated ones. Each of these sub-fields can be further subdivided based on the choice of representation of the fat jets. While initial jet-tagging approaches heavily relied on carefully constructed High-Level Features (HLFs, the Jet Substructure Observables)\cite{Datta:2017rhs,Aguilar-Saavedra:2017rzt,Luo:2017ncs,Moore:2018lsr,Datta:2017lxt,Komiske:2017aww}, modern deep-learning approaches \cite{deOliveira:2015xxd,Baldi:2016fql,Komiske:2016rsd,ATL-PHYS-PUB-2017-017,Kasieczka:2017nvn,Bhimji:2017qvb,Macaluso:2018tck,Guo:2018hbv,Dreyer:2018nbf,Guest:2016iqz,Louppe:2017ipp,Cheng:2017rdo,Egan:2017ojy,Fraser:2018ieu,Almeida:2015jua,Pearkes:2017hku,Butter:2017cot,Roxlo:2018adx, Choi:2018dag, Dreyer:2020brq,Gong:2022lye, Bogatskiy:2020tje, Qu:2019gqs, Moreno:2019bmu, Mikuni:2021pou, Konar:2021zdg, Shimmin:2021pkm,CMS:2020poo, Barnard:2016qma, Lin:2018cin, Du:2019civ, Li:2020grn, Filipek:2021qbe, Bols:2020bkb, Erdmann:2018shi, Moreno:2019neq, Mikuni:2020wpr, Bernreuther:2020vhm, Guo:2020vvt, Dolan:2020qkr} directly utilize the low-level information/features (LLFs) and allow the training process to construct a powerful discriminant for the classification task. However, it is often unclear whether the classifier learns all the relevant information there is. The question also arises whether it is possible to enhance the performance further by incorporating efficient HLFs alongside the LLFs during the network design. To address these issues (and others discussed later), we introduce an architectural modification that can be used with most GNN architectures to include HLFs in the training process as a regularisation mechanism and enhance the overall performance. To demonstrate the method's usefulness, we show its implementation for the case of quark/gluon tagging.\\

Most GNN architectures employed for boosted object tagging \footnote{Note that we are using the term boosted object tagging over fat jet tagging as for the case of quark/gluon tagging, we do not need large radius jets to reconstruct the objects} can be conveniently divided into three blocks. The first is the input block, which processes raw input data into a format suitable for the network. The second is the main building block, where the core transformations are performed, often involving aggregation and message-passing operations that define the network’s design. While the distinction between the input block and the main building block is sometimes blurred, this does not impact the overarching strategy we propose. The final component is the decoding block, where the output from the main block is linearized and passed through fully connected layers. The final layer typically has as many nodes as there are classes, with each node representing a scalar value—a function of the weights and biases of all preceding layers—corresponding to the classification task. We propose an enhancement to this structure by replacing these scalar outputs with vectors. In this scheme, the vector's orientation (its components) encodes multiple characteristics of the object under study, while its magnitude represents the confidence that the jet belongs to a particular class. In other words, instead of a single neuron, we need to associate each node with a bunch of neurons. Such constructions already exist in the literature and are popularly known as capsules. (Note that instead of just the last layer, we will replace the decoding block with capsule layers). \\

Capsule Networks (CapsNets) are first introduced in ref \cite{10.1007/978-3-642-21735-7_6,sabour2017dynamic}, where the authors have used it to get state-of-the-art performance on the MINST dataset \cite{LeCun2005TheMD}. Subsequently, Ref.~\cite{Katebi_2019} applied capsules to galaxy morphology predictions, and Ref.~\cite{Lukic_2019} utilized them for the classification of radio galaxies. In the particle physics community, Ref.~\cite{Diefenbacher:2019ezd} employed capsules for identifying a resonance decaying into a top-quark pair from continuum top and QCD dijet backgrounds. Notably, in all the above cases\footnote{An implementation of capsules with GNNs already exists in the literature and is called CapsGNN~\cite{xinyi2018capsule}. The CapsLorentzNet architecture discussed here bears no resemblance to CapsGNN in terms of its architecture and conceptualization.}, capsules were presented as natural extensions of CNNs and were applied to the image representation of jets. We believe capsules are a powerful tool and should not be restricted to CNNs only. For our analysis, we propose a version of capsules designed for use with GNNs, aiming to enhance their performance. While the capsule structure itself remains unchanged, the manner in which inputs are provided to the capsules is adapted for GNNs. Capsule networks possess a critical feature: they use reconstruction as a regularization method. We took advantage of this and trained the capsule block to reconstruct many of the High-Level Features (HLFs) important for the quark and gluon jets. Note that these HLFs go as input to the main neural network, but the capsule network only sees the output of the main block and, therefore, does not have direct access to these HLFs. This step will ensure that our NN learns all the relevant information from the input data (See Section \ref{sec:model} for the details). As mentioned, we will only use these capsules in the decoding layer. For the complete architecture, we also need a main network. For this task, we choose to work with LorentzNet \cite{Gong:2022lye}, a Lorentz equivariant, permutation equivariant GNN. Our goal is to demonstrate the performance improvement achieved by replacing the conventional decoding block with a capsule-based decoding block.\\ 

It is important to note that some of the HLFs employed in our analysis (see Section \ref{sec:dataset}) are not Lorentz invariant. Consequently, unlike LorentzNet, the CapsLorentzNet architecture does not preserve Lorentz equivariance. While Lorentz equivariance is undoubtedly an elegant and desirable property, the quark-gluon tagging problem is inherently complex, and our primary focus in this work is on developing an architecture that can improve the performance of existing state-of-the-art models for this task. Our analysis demonstrates that the proposed architecture achieves this objective, albeit at the cost of losing Lorentz equivariance. We argue that this trade-off is both deliberate and justified, as it enables meaningful improvements in model performance, which are paramount in addressing the challenges posed by the quark-gluon tagging problem.\\

The main findings of this article are as follows:
\begin{itemize}
	\item We propose an implementation of capsules with LorentzNet (CapsLorentzNet), which can be extended to any other GNN.
	\item We have demonstrated a technique for efficiently incorporating HLFs into the classification task along with the LLFs. To this end, we have first proved that including HLFs directly in the GNN as additional Global variables produces little improvement. This happens because the network treats these HLFs on the same footing as the LLFs without recognizing their significance. Our method provides a mechanism for the network to pay more attention to these HLFs, allowing it to take maximum advantage of these features.
\end{itemize}

The rest of the paper is organized as follows. In section \ref{sec:dataset}, we discuss the dataset used for our analysis along with the validation of our methodology. In section \ref{sec:model}, we discuss the model architecture. This includes a summary of the LorentzNet architecture together with the details of the capsulated decoding layer. Section \ref{sec:performance} discusses the performance of our architecture for quark-gluon tagging. Finally, in section \ref{sec: summary}, we conclude our discussion.

\section{Dataset}
\label{sec:dataset}
To evaluate the model's performance, we selected the task of quark versus gluon jet discrimination, a widely used benchmark in particle physics. Quark-gluon tagging has been extensively studied, and many prior models have validated their performance using the publicly available dataset~\cite{komiske_2019_3164691}, originally introduced in Ref.~\cite{Komiske:2018cqr}. For our analysis, we generated a dataset following the methodology described in Ref.~\cite{Komiske:2018cqr}. 


The dataset comprises one million quark jets and one million gluon jets. These parent processes were generated using \textsc{Pythia8}~\cite{Bierlich:2022pfr}. Quark jets were produced from the process $qg \rightarrow \gamma/Z q$, while gluon jets were generated via $q\bar{q} \rightarrow \gamma/Z g$. In both cases, photon processes were ignored, and the final-state $Z$ boson was decayed into neutrinos. The light flavor quarks ($u$, $d$, $s$, and their antiparticles) were used in the classification task. The center-of-mass energy was fixed at $\sqrt{s} = 14~\mathrm{TeV}$. Initial state radiation (ISR), final state radiation (FSR), and multi-parton interactions (MPI) were enabled during the simulation.  Final-state visible particles were clustered into $R=0.4$ anti-$k_t$ \cite{Anti-KT} jets using \textsc{FastJet3}~\cite{Cacciari:2011ma}. Jets were required to lie within a transverse momentum range of $500~\mathrm{GeV} < p_T < 550~\mathrm{GeV}$ and satisfy $|\eta| < 1.7$. For the final analysis, the four-momenta of the jet constituents and their corresponding PDG IDs were stored.

For our analysis, in addition to the abovementioned information, we need information on some high-level features (HLFs) that can effectively discriminate quark-initiated jets from gluon-initiated jets. For our analysis, we considered ten such HLFs (Most of these HLFs are discussed in \cite{Gallicchio:2011xq, Krohn:2012fg, CMS:2013kfa, Larkoski:2013eya, Pumplin:1991kc, ATLAS:2016wzt}. We urge the interested reader to consult these references for more details.). The list includes:
\begin{itemize}
	\item The invariant mass
	\item The jet charge
	\item The ratio of subjettiness variables \cite{Thaler:2010tr} : $\tau_{21}$, $\tau_{32}$, and $\tau_{43}$.
	\item Number of charged tracks inside the jet
	\item Track based $p_T$ weighted width of the jet: 
	\begin{equation}
		\frac{\sum_i p_{Ti} \Delta R_{i,J}}{\sum_i p_{Ti}}
	\end{equation}
	Where the sum runs over all the tracks inside the jet, and J stands for the jet.
	\item Overall $p_T$ weighted width of the jet:
	\begin{equation}
		\frac{\sum_i p_{Ti} \Delta R_{i,J}}{\sum_i p_{Ti}}
	\end{equation}
	The sum runs over all the constituents of the jet, and J stands for the jet.
	\item The fraction of $p_T$ carried by the highest $p_T$ constituent.
	\item The two-point energy correlation function:
	\begin{equation}
		\frac{\sum_{i, j} E_{Ti}E_{Tj} \Delta R_{i,j}^{\beta}}{\left(\sum_i E_{T,i}\right)^2}
	\end{equation}
	Here, the indices i and j run over all the jet constituents, and we have fixed $\beta$ at 0.2.
\end{itemize}

\begin{figure}[!htb]
	\centering
	\includegraphics[width=0.8\columnwidth]{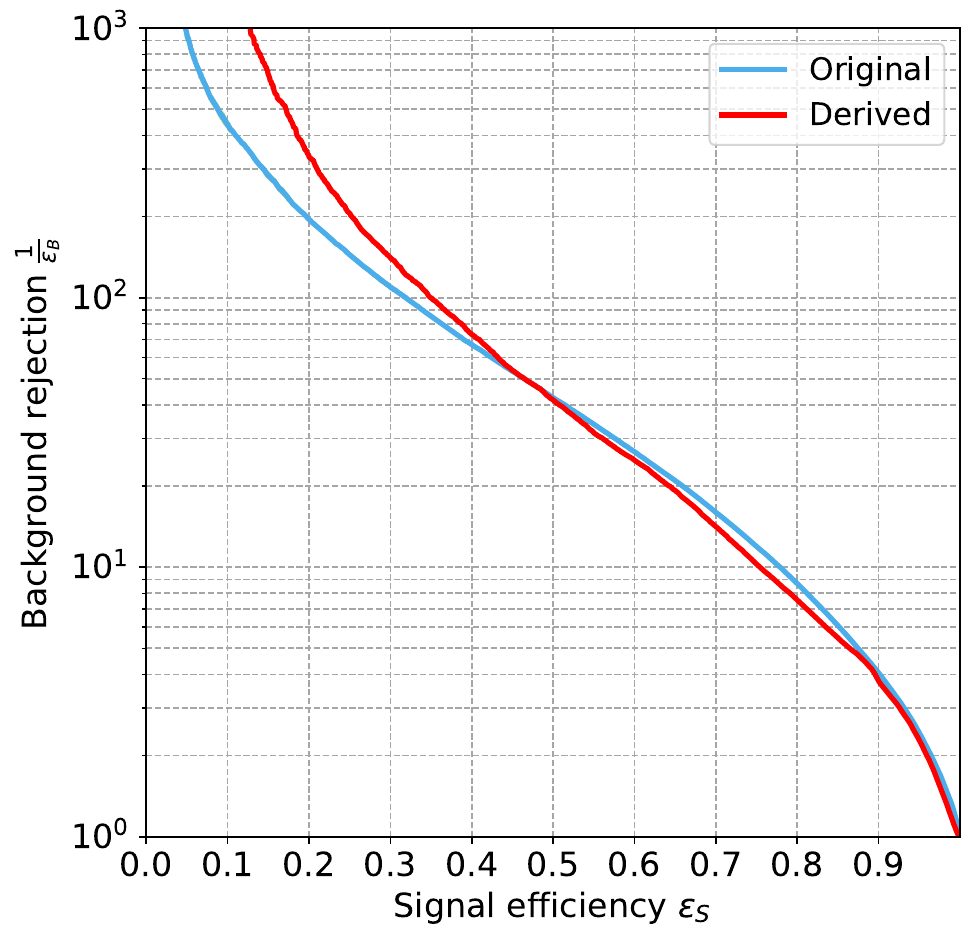}
	\caption{ROC curves of the LorentzNet classifier trained with the publicly available dataset \cite{komiske_2019_3164691} (blue) and our generated dataset (red).}
	\label{fig:val}
\end{figure}

Before proceeding with our analysis, ensuring the correctness of our methodology is crucial. For this, we train the LorentzNet \cite{Gong:2022lye} model on both the publicly available datasets and our generated dataset and compare its performance to test the similarity between the generated and preexisting datasets. We present our results in Figure \ref{fig:val}. As evident from the figure, the ROC curves for the two datasets match quite well for higher signal efficiencies. The greater discrepancy for smaller (larger) signal efficiency (background rejection) can be ascribed to the uncertainty of the data generation process. Since we are not comparing the performance of our architecture with any of the publicly available models, our dataset is self-sufficient, and the higher discrepancy for smaller signal efficiency with the publicly available dataset is not an issue. We have also presented the performance of LorentzNet with the two datasets in Table \ref{tab:val}.

\begin{table}[htb!]
	\centering
	\begin{tabular}{l@{\hspace{2mm}}|l@{\hspace{2mm}}|l@{\hspace{2mm}}|l@{\hspace{2mm}}|r}
		\toprule
		Dataset   &   Accuracy    &   AUC   &   $1/\epsilon_B (\epsilon_s = 0.5)$   &   $1/\epsilon_B(\epsilon_s = 0.7)$ \\
		\midrule
		Original   &  0.8439 & 0.916  & 42.57 $\pm$ 0.18 & 15.97 $\pm$ 0.12    \\
		Derived    &  0.834  & 0.9089 &      41.79       &   14.14   \\ 
		\bottomrule
	\end{tabular}
	\caption{Performance comparison between LorentzNet trained with the publicly available dataset and our generated dataset.}
	\label{tab:val}
\end{table}

\section{Model}
\label{sec:model}
As outlined in the introduction, we aim to demonstrate the effectiveness of capsules in enhancing the performance of GNNs. To serve our purpose, we choose to work with the LorentzNet architecture \cite{Gong:2022lye}, augmenting it with an additional capsule block as the decoding block of the network. In the following, we will briefly review LorentzNet and the capsule block. \\

\subsection{The LorentzNet block}
\label{lorentznet}

LorentzNet is a permutation equivariant and Lorentz equivariant Graph Neural Network (GNN). It follows the universal approximation theorem \cite{villar2023scalars} to construct Lorentz group equivariant continuous mappings. LorentzNet takes as input the jet constituents' four vectors ($x_i$) and the associated scalars ($h_i$). In the original architecture, these scalars were constructed by one-hot encoding of the particle identification (PID). For our analysis, we have also included ten additional HLFs (See section \ref{sec:dataset}) as global variables. These inputs are made to pass through an input layer that projects these scalars into a higher dimensional latent space. \\

The building block of LorentzNet is called the Lorentz Group Equivariant Block (LGEB). It receives the output of the input layer and processes it further. The main action of the LGEBs can be summarised in three steps. Firstly, it utilizes the node embedding scalars, the Minkowski dot products, and the Minkowski norms of the jet constituent four vectors to construct the edge functions between nodes. Next, it uses these edge functions to construct the Minkowski dot product attention, aggregating the neighborhood four vectors and updating the node coordinates. Finally, it aggregates the neighborhood edge functions by weighting them with an edge significance measure and updates the node embedding scalars. In LorentzNet, several LGEBs are used in series. If we denote the output of the $l$th LGEB as ($x^l,h^l$), then the action of the $(l+1)$th LGEB can be summarised as \cite{Gong:2022lye}:
\begin{align}
	m_{ij}^l &= \phi_e\left( h_i^l,h_j^l,\psi(||x_i^l-x_j^l||^2,\psi(\langle x_i^l,x_j^l\rangle))\right)\\
	x_i^{l+1} &= x_i^l + c \sum_{j \in [N]} \phi_x(m_{ij}^l).x_j\\
	h_i^{l+1} &= h_i^l + \phi_h\left( h_i^l, \sum_{j \in [N]} w_{ij} m_{ij}^l\right)
\end{align}
where, $w_{ij} = \phi_m(m_{ij}^l)$ characterize the significance of the neighboring edges, ($\phi_x,\phi_e,\phi_h,\phi_m$) are neural networks, $\psi(x) = sign(x)log(|x|+1)$ is a normalizing function and c is a constant introduced to prevent the scale of $x_i$ from exploding. In the original paper, the model was constructed with six LGEBs. We found five such blocks are enough for our study. We encourage the interested reader to consult ref \cite{Gong:2022lye} for details on the architecture.\\

In the original LorentzNet model, the output of the LGEBs is passed through a decoding layer where the node embeddings ($h$'s) of the final LGEB undergo a global average pooling followed by two fully connected layers with dropout and a final softmax function to generate the classification prediction. For our analysis, we completely removed this decoding layer, replacing it with a fully connected layer. The output of this fully connected layer then passes through the capsule block to generate model predictions.

\subsection{The Capsule block}
\label{capsule}

Capsule Network (CapsNet) \cite{sabour2017dynamic} was originally introduced as an upgraded version of CNNs for image analysis. They have proved quite effective in the analysis of overlapping images. However, as obvious from the previous discussion, we propose using capsules alongside GNNs for object-level prediction tasks. Below, we provide a brief overview of capsules and the design details of the capsule block used in our analysis.\\

The key feature of capsules is that they are vector neurons where the length of the capsule gives the probability of the presence of a given object. At the same time, its orientation represents the different features of the object. A dynamic routing mechanism ensures that the output of the capsule in the lth layer is sent to the appropriate parent in the (l+1)th layer. Say, $u_i$ denotes the output of capsule $i$ in the lth layer. This output is multiplied with a learnable weight matrix to produce the prediction vectors $\hat{u}_{j|i}$ for the jth capsule in the (l+1)th layer. i.e., \cite{sabour2017dynamic}
\begin{equation}
	\hat{u}_{j|i} = W_{ij}u_i
\end{equation}
A weighted sum is performed on the prediction vectors to determine the input $s_j$ of the next layer capsules \cite{sabour2017dynamic}.
\begin{equation}
	s_j = \sum_i c_{ij} \hat{u}_{j|i}
\end{equation}
Here $c$s are the coupling coefficients that sum to one between a capsule $i$ in the lth layer and all capsules $j$ in the (l+1)th layer, i.e., $\sum_j c_{ij} = 1$. This is ensured by expressing these coefficients as $Softmax(b_{ij})$, where $b_{ij}$s are determined following an iterative procedure. Initially, the logits $b_{ij}$ are initiated to zero so that each capsule in the lth layer is connected with all capsules in the next layer with equal weightage. with these initial logits, the initial coupling coefficient $c$'s and (l+1)th layer input $s$'s are computed. These $s_j$ undergo a squashing procedure to calculate the initial output $v_j$ of the jth capsule in the (l+1)th layer \cite{sabour2017dynamic}.
\begin{equation}
	v_j = \frac{||s_j||^2}{1+||s_j||^2} \frac{s_j}{||s_j||}
\end{equation}
This ensures that capsules with small lengths shrink to zero and longer capsules attain a length closer to unity. These output $v_j$'s are used to update the initial logits $b_{ij}$s by encoding the agreement between the output of the lth layer $\hat{u}_{j|i}$ and the output of the (l+1)th layer $v_j$. i.e., \cite{sabour2017dynamic}
\begin{equation}
	b_{ij} = b_{ij} + \hat{u}_{j|i} . v_j
\end{equation}
In other words, the coupling of parent capsule $j$ with the capsule $i$ in the lth layer increases when this dot product is large. This procedure is repeated a predetermined number of times, and the final output $v$s are used for further analysis.\\

As mentioned at the beginning of this section, the length of the capsule represents the probability that the input belongs to a given class. This is ensured by an appropriate choice of the loss function. For each class $k$, the margin loss has the form \cite{sabour2017dynamic}:
\begin{equation}
	L_k = T_k max(0,m^+ - ||v_k||)^2 + \lambda (1-T_k) max(0,||v_k||-m^-)^2
\end{equation}
where $\lambda$ is a weighting parameter. we follow the original work \cite{sabour2017dynamic} and choose $m^+ = 0.9$, $m^- = 0.1$, and $\lambda = 0.5$. In addition to the margin loss, a reconstruction loss term introduces additional regularisation into the training process. In the original work, the output capsules were decoded to reproduce the original image, and the reconstruction loss quantifies the discrepancy between the original and reconstructed image. For our analysis, we use the jet-level high-level features as the analog of the image. The decoding layer of the capsule block tries to reconstruct these jet-level features and matches them with the input values. This step ensures that the network learns these important characteristics of the jet, enhancing its performance. \\

The implementation of the capsule block is similar to that suggested in reference \cite{sabour2017dynamic}. The major difference is in the input to the capsule block. The original architecture used CNNs to generate this input; hence, the capsules' inputs are images represented in the latent space. The capsule layers in reference \cite{sabour2017dynamic} use convolution operations to process these images. In our case, the output of LorentzNet can be viewed as a linear array of numbers. To handle these, we have replaced the convolution layers of reference \cite{sabour2017dynamic} with fully connected layers. The rest of the architecture is almost similar to reference \cite{sabour2017dynamic}. We have one primary capsule layer accepting the output of LorentzNet. The output of LorentzNet has a shape $102\times 72$, where the number 102 corresponds to the jet constituents, and 72 represents the number of hidden dimensions. The action of the primary capsule layer results in $32 \times 36$ 8-dimensional capsules. To achieve this, it first uses a fully connected layer that converts $102\times 72 \rightarrow 102 \times 36$. After this operation, it uses eight fully connected layers for the transformation $102 \times 36 \rightarrow 32 \times 36$ and combines all these outputs. These are the $u_i$s mentioned earlier. The secondary capsule layer has two capsules (corresponding to the signal and background classes) of dimension eight. These capsules receive input from all capsules from the primary layer. They utilize the iterative routing mechanism discussed earlier to generate the output $v_j$s. Finally, we have the decoding layer that uses the output of the secondary capsules to reconstruct the jet-level features. We implement this decoding layer using three fully connected neural networks. We have made the model and our results available in the public repository: \href{https://github.com/rama726/CapsLorentzNet.git}{CapsLorentzNet}. The interested reader is encouraged to consult the code for fine details regarding the architecture.

We implemented the CapsLorentzNet with PyTorch and trained it on a cluster with four Nvidia Tesla K80 GPUs. We pass the data in batches of size 32 on each GPU. The model is trained for a total of 35 epochs. At the end of each epoch, we test the model performance with the validation dataset, and the one with the best validation accuracy is saved for testing. For the optimizer and learning rate scheduler, we have followed the prescription of reference \cite{Gong:2022lye}. We have used the ADAMW \cite{loshchilov2019decoupled} optimizer with a weighted decay of 0.01. For the details on the learning rate scheduler, see ref \cite{Gong:2022lye}.

\section{Classifier Performance}
\label{sec:performance}
This section will focus on discussing the performance of the CapsLorentzNet architecture. However, before discussing our final results, it will be informative to check how much gain in performance can be achieved if we include these jet-level features (See Section \ref{sec:dataset}) in the LorenzNet itself with small architectural changes. For this purpose, we employed two distinct strategies for integrating these high-level features (HLFs) into the network. In the first approach, referred to as LorentzNet-G1, the HLFs are used as additional node-level features. Specifically, these features are treated as global node attributes shared across all nodes in a graph and combined with the constituent-level information (e.g., PIDs) to form the node inputs. In this way, the HLFs provide global characteristics that complement the local constituent-level information. In the second approach, termed LorentzNet-G2, the HLFs are introduced at the decoding block of LorentzNet. Here, the HLFs are combined with the processed constituent-level information after passing through the network's main blocks. This approach aims to use the HLFs to refine the final predictions directly. For LorentzNet-G1, we have changed the size of the input layer of LorentzNet to accommodate the additional HLFs. Similarly, for LorentzNet-G2, we have changed the size of the first layer of the decoding block of LorentzNet. \\

\begin{figure}[!htb]
	\centering
	\includegraphics[width=0.8\columnwidth]{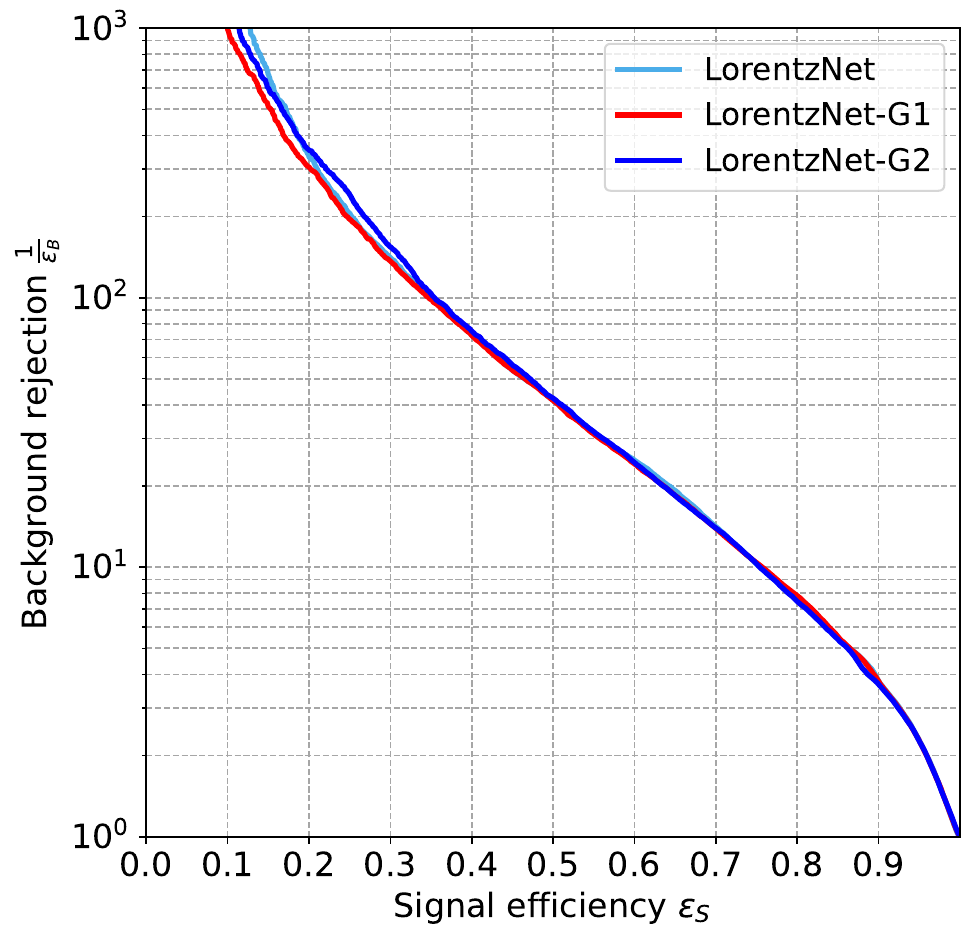}
	\caption{ROC curves of the LorentzNet classifier without jet-level information (cyan) and the two classifiers with additional global information (red and blue).}
	\label{fig:LNglob}
\end{figure}

The results of these modifications are summarized in Figure~\ref{fig:LNglob} and Table~\ref{tab:LNglob}. As evident from the results, incorporating these additional jet-level features does not yield a significant improvement in the classifier's performance. For LorentzNet-G1, the lack of improvement likely arises because the classifier is treating these jet-level features on the same footing as the PID information of the constituents without understanding their importance for the classification task at hand. Similarly, the results for LorentzNet-G2 highlight the inherent complexity of the quark-gluon tagging problem and suggest that these straightforward modifications to the architecture are insufficient to enhance its performance meaningfully.\\

\begin{table}[htb!]
	\centering
	\begin{tabular}{l@{\hspace{2mm}}|c@{\hspace{2mm}}|c@{\hspace{2mm}}|c@{\hspace{2mm}}|c}
		\toprule
		Model   &   Accuracy    &   AUC   &   $1/\epsilon_B (\epsilon_s = 0.5)$   &   $1/\epsilon_B(\epsilon_s = 0.7)$ \\
		\midrule
		LorentzNet    &  0.834  & 0.9089 &  41.79   & 14.14   \\ 
		LorentzNet-G1 & 0.837 & 0.9087 & 41.55 & 13.91    \\
		LorentzNet-G2 & 0.8334 & 0.9080 & 42.34 & 13.97    \\
		\bottomrule
	\end{tabular}
	\caption{Performance comparison between LorentzNet trained with PID information of jet constituents as node embeddings, LorentzNet-G1 that includes additional jet-level observables in the node embeddings, and LorentzNet-G2 where these global information are used directly at the decoding block (see text for details).}
	\label{tab:LNglob}
\end{table}

\begin{figure}[!htb]
	\centering
	\includegraphics[width=0.45\columnwidth]{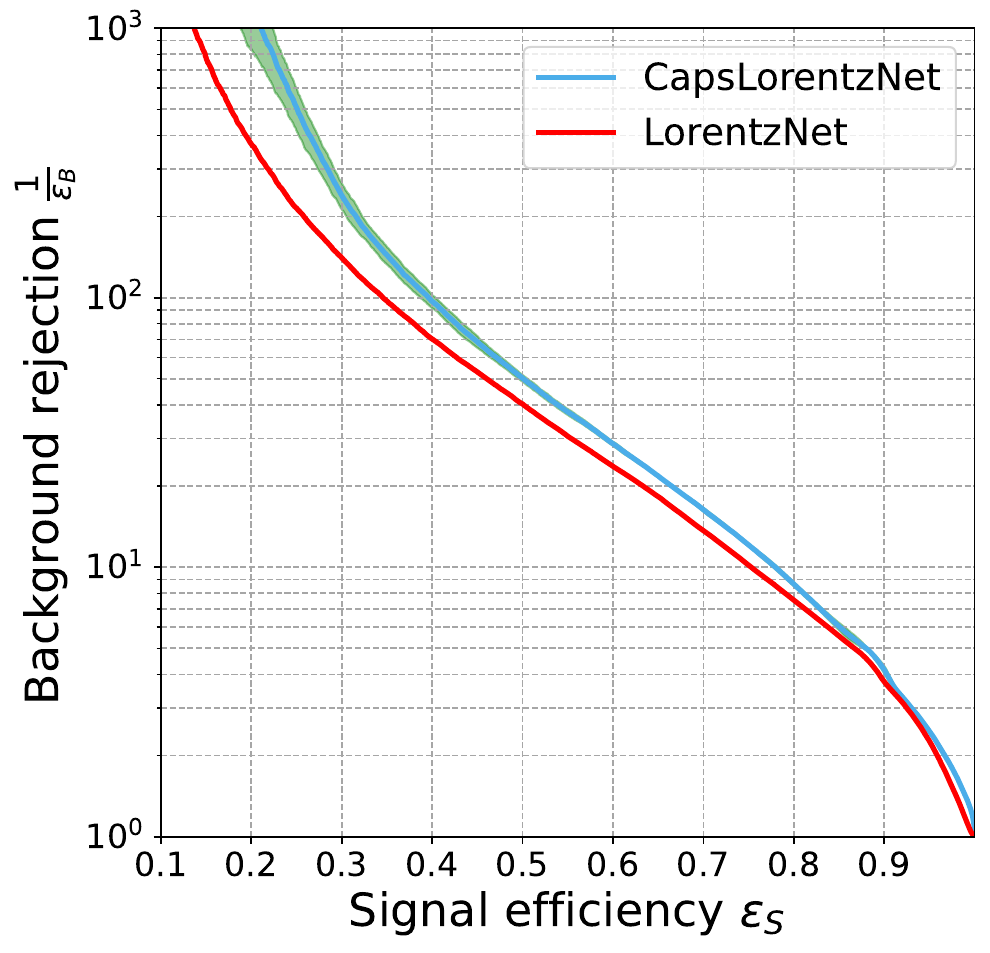}
	\includegraphics[width=0.45\columnwidth]{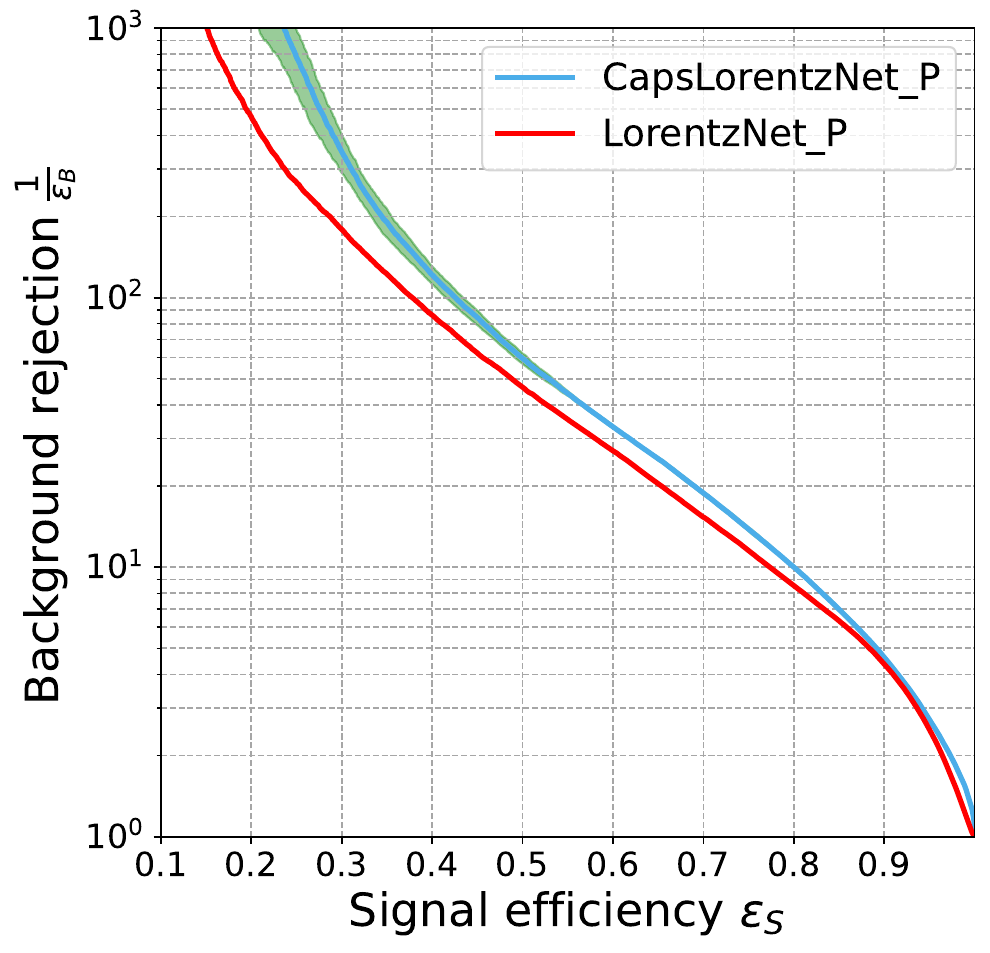}
	\caption{ROC curves of the LorentzNet (red) and CapsLorentzNet (blue) Classifiers. The left plot depicts the ROC curves for the models when both training and testing are performed on our dataset. Conversely, the right plot illustrates the performance of the models trained on our dataset but evaluated on a public dataset. The results are averaged over six runs with different random seed initializations.}
	\label{fig:final}
\end{figure}

Having established the necessity of the CapsLorentzNet architecture, we now present its performance on the quark-gluon dataset. In Figure~\ref{fig:final} (left panel), we show the ROC curve for the CapsLorentzNet classifier. For comparison, the ROC curve for the original LorentzNet architecture is also included. Both results are averaged over six runs with different random seed initializations to account for variability. For a consistent comparison, key performance metrics are summarized in Table~\ref{tab:final} (first and second rows).\\

Notably, a comparison of background rejection at 50\% signal efficiency reveals that CapsLorentzNet achieves an approximately 20\% performance gain over LorentzNet. While this improvement might not appear striking at first glance, it is worth highlighting that the state-of-the-art architecture for quark-gluon tagging, ParT~\cite{Qu:2022mxj}, achieves a background rejection of $47.9 \pm 0.5$ at 50\% signal efficiency. However, it should be noted that the datasets used to train ParT and CapsLorentzNet, although similarly generated, are not identical. A fair, direct comparison would require both models to be trained and tested on the same dataset.\\

To ensure completeness, we also evaluate the performance of the trained LorentzNet and CapsLorentzNet models on the public quark-gluon dataset~\cite{komiske_2019_3164691}. These results are summarized in the right panel of Figure~\ref{fig:final} and in the last two rows of Table~\ref{tab:final}. To avoid confusion, we label the results obtained from testing our models on the public dataset as LorentzNet\_P and CapsLorentzNet\_P.\\

CapsLorentzNet consistently outperforms LorentzNet, achieving a 20–25\% improvement in performance, as demonstrated across both our generated (first two rows of Table \ref{tab:final}) and the public dataset (last two rows of Table \ref{tab:final}). This robust improvement indicates that the observed gains are intrinsic to the model and not merely a result of dataset-specific biases. While we observe an overall performance shift for both models when tested on the public dataset, this shift does not reflect network uncertainty. Instead, it arises from differences in the datasets used for testing the classifiers: our training/testing dataset was generated with \textit{Pythia 8.306}, whereas the public dataset was generated with \textit{Pythia 8.226}. Additionally, methodological differences between our approach and that of the public dataset authors may have contributed to this variation. Despite these factors, the consistent relative performance of CapsLorentzNet highlights its generalizability and resilience across datasets.

For reproducibility, we have made our code and dataset publicly available on GitHub: \href{https://github.com/rama726/CapsLorentzNet.git}{CapsLorentzNet}.

\begin{table}[htb!]
	\centering
	\begin{tabular}{l@{\hspace{2mm}}|c@{\hspace{2mm}}|c@{\hspace{2mm}}|c@{\hspace{2mm}}|c}
		\toprule
		Model   &   Accuracy    &   AUC   &   $1/\epsilon_B (\epsilon_s = 0.5)$   &   $1/\epsilon_B(\epsilon_s = 0.7)$ \\
		\midrule
		LorentzNet    &  0.834 $\pm$ 0.0004  & 0.9089 $\pm$ 0.0008 &  40.34 $\pm$ 0.97   & 13.67 $\pm$ 0.28   \\ 
		CapsLorentzNet & 0.843 $\pm$ 0.0009& 0.91987 $\pm$ 0.0011 & 50.26 $\pm$  1.43& 16.36 $\pm$ 0.27     \\
		LorentzNet\_P    &  0.845 $\pm$ 0.001  & 0.917 $\pm$ 0.0006 &  46.58 $\pm$ 1.05   & 15.34 $\pm$ 0.29   \\ 
		CapsLorentzNet\_P & 0.853 $\pm$ 0.0009& 0.928 $\pm$ 0.0009 & 59.52 $\pm$  2.58 & 18.85 $\pm$ 0.29     \\
		\bottomrule
	\end{tabular}
	\caption{Performance comparison between LorentzNet and CapsLorentzNet. The first two rows summarize the performance of the models when both training and testing are conducted on our dataset. In contrast, the last two rows present results where the models are trained on our dataset but tested on a public dataset. All results are averaged over six runs with different random seed initializations.}
	\label{tab:final}
\end{table}

\section{Summary and Outlook}
\label{sec: summary}
This article proposes a new and innovative architectural modification that can be incorporated into most GNN architectures. It advocates replacing the final fully connected layers of any GNN architecture (decoding block) with capsules. This allows the scalar activations of the decoding blocks to be replaced with vector activation of the capsules. The orientation of these vectors represents important characteristics of the objects under study, and the magnitude characterizes whether the object belongs to the class the capsule represents. The regularisation by reconstruction mechanism of the capsule network provides a means to consistently incorporate High-Level Features into the analysis along with the usual low-level inputs of the GNNs. We have demonstrated the effectiveness of this strategy with the LorentzNet architecture for quark-gluon tagging. Our analysis shows a performance gain of around 20 \% can be achieved with this strategy. This novel architectural modification can be extended to any other GNN architecture, and we believe similar performance gains can be achieved in all of these cases.

\section*{ACKNOWLEDGMENTS}
RS thanks Kirtiman Ghosh for insightful and helpful discussions during the preparation of this manuscript. The simulations were partly supported by the SAMKHYA: High-Performance Computing Facility provided by the Institute of Physics, Bhubaneswar.

\newpage
\bibliographystyle{JHEP}
\bibliography{v1}

\end{document}